\documentclass{article}
\usepackage{authblk}

\usepackage{blindtext}
\usepackage{graphicx}
\usepackage[utf8]{inputenc}
\usepackage{graphicx}
\usepackage{blindtext}
\usepackage{amsmath,amsfonts,amssymb}
\usepackage{mathtools}
\usepackage[]{algorithm2e}
\usepackage{titlesec}

\title{Analysis of vocal breath sounds before and after administering Bronchodilator in Asthmatic patients}

\author{Shivani Yadav$^1$, Dipanjan Gope$^2$, Uma Maheswari K.$^3$, \\Prasanta Kumar Ghosh$^4$}
\affil{$^1$BioSystems Science and Engineering, Indian Institute of Science (IISc), Bangalore-560012, India\\
	$^2$Electrical Communication Engineering, Indian Institute of Science (IISc), Bangalore-560012, India\\
	$^3$Pulmonary Medicine, St. Johns National Academy of Health Sciences, Bangalore-560034, India\\
	$^4$Electrical Engineering, Indian Institute of Science (IISc), Bangalore-560012, India}

\date{}

\begin{document}
\maketitle
\section*{Abstract}
Asthma is one of the chronic inflammatory diseases of the airways, which causes chest tightness, wheezing, breathlessness, and cough. Spirometry is an effort dependent test used to monitor and diagnose lung conditions like Asthma. Vocal breath sound (VBS) based analysis can be an alternative to spirometry as VBS characteristics change depending on the lung condition. VBS test consumes less time, and it also requires less effort, unlike spirometry. In this work, VBS characteristics are analyzed before and after administering bronchodilator in a subject-dependent manner using linear discriminant analysis (LDA). We find that features learned through LDA show a significant difference between VBS recorded before and after administering bronchodilator in all 30 subjects considered in this work, whereas the baseline features could achieve a significant difference between VBS only for 26 subjects. We also observe that all frequency ranges do not contribute equally to the discrimination between pre and post bronchodilator conditions. From experiments, we find that two frequency ranges, namely 400-500Hz and 1480-1900Hz, maximally contribute to the discrimination of all the subjects. The study presented in this paper analyzes the pre and post bronchodilator effect on the inhalation sound recorded at the mouth in subject dependent manner. Findings of this work suggests that, inhalation sound recorded at mouth can be a good stimulus to discriminate pre and post bronchodilator conditions in asthmatic subjects. Inhale sound based pre and post bronchodilator discrimination can be of potential use in clinical settings.
\section*{Keywords}
	Asthma, Bronchodilator, Breath sounds, Spirometry, Linear discriminant analysis, Inhale

\section{Introduction}

Asthma is a chronic inflammatory disease of the airways, which causes 1000 deaths every day around the world \cite{networkglobal}. Symptoms of asthma includes breathlessness, chest discomfort, cough, wheeze, and other peculiar sounds during breathing \cite{martinez2007genes}.
For asthma treatment, bronchodilators, such as beta-2 agonists, anticholinergics, and theophylline, are used regularly. Bronchodilators reduce the inflammation and mucous secretion in the airways that makes breathing easier for an asthmatic patient. 
Bronchodilators are generally administered through the mouth with the help of a device called an inhaler.
Lung function tests are used to monitor and diagnose asthma by measuring lung capacity, lung volume, etc. Spirometry is one of the gold standard lung function tests. 
Spirometry measures the expelled volume and flow of air from the lungs.
Three variables, namely, Forced vital capacity (FVC), Forced expiratory volume in one second (FEV1), and the ratio, of FEV1/FVC, are typically measured by spirometry. Pulmonologists compare FVC, FEV1, and FEV1$/$FVC before and after taking bronchodilator diagnose asthma and its severity and to rule out the possibility of having other obstructive diseases including Chronic obstructive pulmonary disease (COPD). 




Spirometry is an effort-dependent and time-consuming test as it requires continuous guidance by a technician. Such a long and arduous process makes testing very difficult, especially for the children and older people \cite{national1998expert}. Peak Flow Meter (PFM) is another technique used for home monitoring, and ambulatory evaluation of asthma \cite{devrieze2017peak}. Limitations of the peak flow meter include the fact that it measures the peak expiratory flow rate through major airways but fails to measure the flow rate through minor airways, which can also be affected during asthma. PFM and spirometry readings are also affected by several factors like inadequate efforts, lips not tight around the mouthpiece, tongue blocking mouthpiece. \cite{Dowd}. Hence, it is required to have a technique to monitor asthma that can overcome these limitations at home and in clinical settings. The sound-based technique can be one of the alternatives for the task \cite{jane1998spectral}. Dogan et al. \cite{dogan2007subjective} have shown voice quality parameters like maximum phonation time, frequency, and amplitude perturbation are shown to be impaired in asthmatic subjects.

Breath sounds recorded at the chest and the mouth have been studied to determine the lung condition. Breathing sounds recorded at the anterior chest, posterior chest, and trachea with stethoscopes or microphones are referred to as breath sounds. On the other hand, breathing sounds recorded at the mouth are referred as vocal breath sounds (VBS) in this work. It has been reported in the literature that airways obstruction changes breath sound characteristics. 
Pardee et al. \cite{pardee1976test} have observed a strong correlation between the loudness of lung sounds and FEV1 value in a study using 183 patients. They have concluded that breath sounds intensity alters with obstructive pulmonary diseases. Breath sounds may show some changes even in the absence of adventitious sounds like wheeze \cite{malmberg1994frequency}. Wheeze is a whistling sound produced during breathing due to the obstruction in the airways \cite{forgacs1971breath}. Wheezing can be observed during inhalation, exhalation, or both \cite{pramono2017automatic}. Spontaneous wheezing is found to be present during inspiration in adults \cite{shim1983relationship} and children \cite{sanchez1993acoustic} suffering from asthma. Many works have been reported in the past to monitor or diagnose asthma using wheeze sound recorded at the chest. Several methods have been reported in the literature for wheeze detection, including Gaussian mixture models with sub-band based cepstral parameters \cite{bahoura2004respiratory}, Welch spectrum with feed-forward neural networks \cite{oud2003lung}, and time-frequency spectrum \cite{forkheim1995comparison}. Some methods used the spectral and temporal characteristics of breath sounds to assess the effect of bronchodilators.
Tabata et al. \cite{tabata2018changes} showed that the spectral characteristics of breath sounds before and after going through methacholine challenge test, followed by bronchodilation, are significantly different. Malmberg et al. \cite{malmberg1994frequency} also studied the relationship between breath sounds median frequency and FEV1 in histamine challenge test and after subsequent bronchodilation in asthmatic patients. They have found a significant increase in the median frequency of breath sound spectra after histamine inhalation in asthmatic subjects.  
Jane et al. \cite{jane1998spectral} have analyzed central frequency 0-2500Hz, modified central frequency between 300-2500Hz range, and power in the signal above 300Hz in tracheal sounds, to assess the effect of bronchodilation in asthmatic and healthy subjects. Central frequency is defined as the frequency at which power of the signal becomes 50\% of the total power of the signal. They have observed that the central frequency between 0-2500Hz has shown a significant decrease after administering the bronchodilator in the asthmatic group, whereas there was no significant change in the healthy group.
Along with breath sounds, cough sounds have also been used to find changes in their characteristics before and after bronchodilator. Cough is produced by expiratory muscles contraction against a closed glottis and a sudden release of pressure afterward \cite{evans1975mechanical}. Thrope et al. \cite{thorpe2001acoustic} have reported temporal envelope, power spectrum, zero crossings, which show significant changes before and after bronchodilation challenge test in 20 subjects.

Part of our research interests is voice-based monitoring and diagnosis of asthma, as such a technique helps patients irrespective of age and medical conditions and requires less clinical training.
To the best of our knowlwdge, very few approaches have used VBS for monitoring and diagnosing asthma. 
Forgacs et al. \cite{forgacs1971breath} have analyzed the inhalation sound recorded at the mouth in chronic bronchitis and asthmatic patients. They noted that intensity of inhale signal is higher in patients compared to healthy subjects at identical flow rates.  They hypothesized that loud inhalation breath sounds are generated because of the turbulent airflow caused by the narrowing of bronchi and its segmental and lobar regions. They have also observed a reduction in the intensity of inhalation breath sounds after administering bronchodilators. This called for an in-depth analysis of the inhale phase of VBS (here on referred as IPVBS) before and after taking bronchodilator, which is addressed in this work. We hypothesize that the sound generated at mouth depends on the lungs volume \cite{iwarsson1998effects}. Hence, if a pathology affects lung volume, it would reflect in the patient's voice. 
In our previous work \cite{yadav2018comparison}, we have compared cough, VBS, and sustained phonations recorded at mouth of 35 patients and 36 healthy for asthma versus healthy subjects classification. VBS is found to provide the highest classification accuracy of 89.8\% among all the sounds.

In this study, subject-dependent spectral analysis of IPVBS before and after administering a bronchodilator has been done to understand the effects of the change in the airway obstruction on IPVBS. We choose a subject dependent setup for IPVBS because it is already known in the literature that no two individuals can have identical vocal tract length, vocal tract shape, and parts of their voice production system \cite{kinnunen2010overview}. Lung function also depends on gender \cite{gross2000relationship} as well as age \cite{john1986longitudinal} \cite{turner1968elasticity}. Thus, subject-dependent analysis reduces the variation introduced by variables such as vocal tract length, age, gender etc., and helps capture changes in VBS characteristics due to airway obstruction and bronchodilation. To perform the spectral analysis, features proposed by Tabata et al. \cite{tabata2018changes} are used as baseline features in this work. We observe that the baseline assumes uniform weights for all the frequency bins to compute the features. In contrast, our proposed features use, a frequency-dependent weighting scheme where frequency dependent weights are learned. Learned weights help understand which frequencies are more useful for discrimination of VBS recorded before and after administering bronchodilators. Weights are learned on the power spectrum of the IPVBS by the using the linear discriminant analysis (LDA) \cite{mika1999fisher}. Data of 30 asthmatic subjects have been used in this study. Statistical testing between features from IPVBS recorded before and after taking bronchodilator is performed using paired t-test \cite{kim2015t} to examine the change in the feature values. Proposed features have shown significant change before and after administering bronchodilator in all 30 subjects. On the other hand, with the best baseline \cite{tabata2018changes} feature, significant change is observed for 26 asthmatic subjects only.
This paper is organized as follows. At first, the data set and description of the proposed approach for LDA based feature learning are explained. Further, experimental setup section describes different kinds of experiments performed in this study. At last, the results and discussion section discusses results of all the experiments carried out in this study, and insights about the results are given.


\section{Materials and Methods}
\subsection{Dataset}
In this study, 30 asthmatic patients' VBS were recorded. Dataset consists of 15 males and 15 females with an average age of 42 years and an age range of 15 to 71 years. The study was conducted under the guidance of a pulmonologist. It was ensured that none of the patients were suffering from another lung disease except asthma. To clinically diagnose asthma, spirometry, as well as patients' history, were considered. Spirometry was performed before and 15 minutes after taking the bronchodilator. The recording was done in the laboratory of St. Johns Medical College Hospital, Bangalore. Consent was taken from each patient before the recordings. The severity of asthma can be mild, moderate, and severe. The severity is decided based on the difference between reference and predicted spirometry variables like FEV1, FVC, and their ratio. The dataset includes 9 mild, 11 moderate, and 10 severe categories of asthmatic subjects. Average FEV1(in litres)(\% predicted)  is $47.66 \pm 18.39$ and average value of FVC(in litres)(\% predicted) is $55.24 \pm 19.03$.

VBS was recorded by using the ZOOM H6 handy recorder at 44100 kHz and 16 bits sampling rate. A microphone was kept in front of the subject's mouth at a distance of approximately 5 cm by the experimenter, who was available throughout the procedure to guide the patients. All recordings were done in the hospital's lung function test laboratory, which has a typical noisy environment with noise sources like airconditioner, fan, and speech babble at the background. During the recording, patients were instructed to be in a sitting position. Throughout the recording, every patient's nose was closed with a nose clip enabling them to exhale up to their full capacity only through the mouth. Patients were instructed to take deep breaths during VBS and not to hurry to complete multiple VBSs. Pre and post-bronchodilator VBS recordings were done after the spirometry. A gap of around 20 minutes was given between pre and post spirometry. Patients were rested for around 10-15 minutes between the two spirometry and the recordings. This is done to avoid fatigue as it can affect breathing, which, in turn, would impact the analysis and conclusions in this study.

The average number of VBSs per subject before and after administering the bronchodilator was eight. 249 exhales and 249 inhales were recorded before administering the bronchodilator, while 238 inhales and 238 exhales after the bronchodilator. Average(standard deviation) duration of VBS recording per subject was 29 ($\pm$ 12.48) secs and 23.38 ($\pm$ 10.55) secs before and after administering the  bronchodilator, respectively. The average (standard deviation) of exhaling duration before and after administering the  bronchodilator was 1.95($\pm$ .75) seconds and 1.61($\pm$ .70) seconds, respectively. Similarly, the mean (standard duration) of inhale before and after administering bronchodilator was 1.48($\pm$ .69) seconds and 1.21 ($\pm$ .49) seconds, respectively. Boundaries of inhale and exhale segments were marked manually by listening, and visual inspection of VBS waveform and spectrogram in Audacity \cite{mazzoni2000audacity}.

\subsection{Analysis of breath sound before and after administering Bronchodilator}

Breath sounds have been shown to be a good predictor of asthmatic condition in a subject, but VBSs are less explored, in the literature, as mentioned by Forgacs et al. \cite{forgacs1971breath}. To increase our understanding of the changes in the spectral properties of IPVBS in asthmatic patients recorded before and after administering bronchodilator, we have analyzed two sets of features. The first set contains baseline spectral features proposed in \cite{tabata2018changes}, and the second set contains features proposed in this work. A description of each set of features is given in the following sub-sections.



\subsubsection{Baseline features}

Baseline features \cite{tabata2018changes} have been analyzed for breath sounds in asthmatic children during methacholine inhalation challenge followed by a bronchodilation test. Spectral features have been calculated by using a high pass filtered inhale sound signal. High pass filtering with a cut-off frequency of 300Hz has been carried out to remove low-frequency puff noise generated when the microphone is kept very close to the mouth. 
Spectral features proposed by \cite{tabata2018changes} have been calculated at the frame level from all inhale sounds in the 300-2000Hz frequency range, after downsampling the signal to 4kHz. Spectral features represent frequency-dependent normalized energy in the signal, defined as the energy ratio in different frequency bands to the signal's total energy and other spectrum-related features. Energy-based features are motivated by the observation that the intensity of inhale signal reduces after bronchodilation, due to reduced turbulent flow \cite{forgacs1971breath}.\\

Although we use the spectral features from work by \cite{tabata2018changes} as the baseline features, there are two key differences between the our work and the that carried out in [17]. Firstly, \cite{tabata2018changes} worked with breath sounds recorded at chest in children, whereas, in this work, those features are used for a study on adults IPVBS.
The frequency range for IPVBS is 200-2000Hz \cite{sarkar2015auscultation}, which is similar to the frequency range of breath sounds analyzed in \cite{tabata2018changes}. Hence, baseline features fits well for the analysis of IPVBS.


The second key difference is that the work in \cite{tabata2018changes} has been done in a subject independent manner. However, a subject dependent analysis is performed in this work. This is mainly because unlike breath sounds at the chest, the VBS is a function of vocal tract shape, articulatory shape and configurations, age and gender of a person, and these change from one person to another.


\subsubsection{Proposed feature}

We hypothesize that not all frequencies in the range of 300Hz-2000Hz will be equally contributing to the discrimination between IPVBS recorded before and after administering bronchodilator. To find out the most discriminating frequencies in this regard, linear discriminant analysis (LDA) is performed on the spectrum of IPVBS. The mathematical formulation of LDA used to compute the proposed feature is briefly described below. Let's say, we have \(N\) frames of IPVBS recorded before bronchodilator and \(K\) frames of IPVBS recorded after administering bronchodilator. Let \(\boldsymbol{x^i_b}\) and \(\boldsymbol{x^j_a}\) denote logarithm of the power spectrum of a $i^{th}$ and $j^{th}$ frame of IPVBS recorded before and after administering bronchodilator, respectively, with \(nf\) number of fast Fourier transform (FFT) bins. Our aim is to find a vector $\boldsymbol{w}$ which can be used to project the frames  \(\boldsymbol{x^i_b}\) and  \(\boldsymbol{x^i_a}\) through the following optimization.
\begin{equation}
\begin{aligned}
\arg\max_{\boldsymbol{w}} \quad  \frac{(\boldsymbol{w^\intercal \mu_b-\boldsymbol{w^\intercal \mu_a)}^2}}{\sigma_{a}^2+\sigma_{b}^2},
\end{aligned}
\end{equation}
where \(\boldsymbol{\mu_b}=\frac{1}{N}\sum_{i=1}^{N}\boldsymbol{x^i_b}\), \(\boldsymbol{\mu_a}=\frac{1}{K}\sum_{j=1}^{K}\boldsymbol{x^j_a}\), \(\sigma_b^2=\sum_{i=1}^{N}(\boldsymbol{w^\intercal x_b^i-w^\intercal \mu_b})^2\) and \(\sigma_a^2=\sum_{j=1}^{K}(\boldsymbol{w^\intercal x_a^j-w^\intercal \mu_a})^2\) are the means and variances of the projected spectrum of the IPVBS frames recorded before and after administering bronchodilator, respectively. The optimization problem in eq. 1 can be rewritten as follows:

\begin{equation}
\begin{aligned}
\arg\max_{\boldsymbol{w}} \quad  \frac{\boldsymbol{w^\intercal}P\boldsymbol{w}}{\boldsymbol{w^\intercal}M\boldsymbol{w}},\\
\end{aligned}
\end{equation}

The solution of eq. 2 can be obtained by solving the generalized eigenvalue problem \cite{ghojogh2019eigenvalue} as follows: 

\[P\boldsymbol{w} = {\lambda}M\boldsymbol{w} \tag{3} \label{eq:eigen},\]

where, \(\lambda\) is an eigen value, \(P\) = \((\boldsymbol{\mu_b-\mu_a})(\boldsymbol{\mu_b-\mu_a})^\intercal\), and \(M\)= \(\sum_{i=1}^{N} (\boldsymbol{x_b^i}-\boldsymbol{\mu_b})(\boldsymbol{x_b^i}-\boldsymbol{\mu_b})^\intercal+\sum_{j=1}^{K} (\boldsymbol{x_a^j}-\boldsymbol{\mu_a})(\boldsymbol{x_a^j}-\boldsymbol{\mu_a})^\intercal
\). 
Dimensions of $P, M$ and $\boldsymbol{w}$ are \(nf\times nf, nf\times nf \) and \( nf\times 1\), respectively.
With the optimum \(\boldsymbol{w}\), spectrum at every frame is projected to a single dimensional feature.
For example, if \(\boldsymbol{f_i}\) is $i^{th}$ frame's spectrum having dimension \( nf\times 1\) and \(\boldsymbol{w_{opt}}\) is the solution of eq. 2, then the proposed 1-dim feature is \(\boldsymbol{w^\intercal f_i}\). To calculate the distance between the distributions of the proposed feature computed from IPVBS recorded before and after administering bronchodilator, the Fisher discriminant ratio (FDR) \cite{kim2006robust} has been used.
\subsection{Experimental setup}

Apart from FDR value, percent change in the mean value of proposed features computed on the VBS recorded before and after taking the bronchodilator has been used. This gives an insight to the amount and direction of change in features' mean values, which is given as follows:
\begin{equation}
\mu_{change\%}=\frac{\mu_a^p-\mu_b^p}{\mu_b^p}\% \quad ,
\end{equation} 
where \(\mu_a^p=\boldsymbol{w_{opt}^\intercal \mu_a}\) 
and \(\mu_b^p=\boldsymbol{w_{opt}^\intercal \mu_b}\) are the averages of the proposed features computed on IPVBS recorded after and before administering the bronchodilator, respectively. Paired t-test \cite{tabata2018changes} has been used to examine the significance in the difference between \(\mu_a^p\) and \(\mu_b^p\). This is done for baseline features too.


VBS recordings are downsampled to 4kHz as it is known that most of the energy lies in a frequency range of 200-2000kHz \cite{sarkar2015auscultation}. Then VBS recordings are high pass filtered using a Butterworth filter of order 6 with a cut-off frequency of 300 Hz. 300 Hz has been chosen to minimize puff noise generated by the sudden burst of air from the mouth during breathing vocally. Features were calculated using 20ms window length and 10ms shift for both the baseline and the proposed features. FFT of order 400 is used for computing the proposed features. However, only 201 points are used due to the symmetry of the spectrum for the analysis. Thus each FFT bin corresponds to 10Hz. However, only bins corresponding to 300-2000Hz are used to calculate the proposed features. Thus, \(nf=171\).

\subsubsection{Baseline features}
Baseline features are calculated following the steps outlined in \cite{tabata2018changes}. 
All 11 baseline features are described briefly here. $P_T$ indicates the total power per frame, $F_{99}$ and $F_{50}$ indicate frequency points which contain \(99 \%\)  and \(50 \%\) energy of the spectrum respectively. $P_2/P_T$ is the ratio of power in the second half of the frequency range considered (i.e., 1150-2000Hz ) and $P_T$. Similarly, $P_3/P_T$ is the ratio of power in the last one-third of the frequency range considered (i.e., 1434-2000Hz) and $P_T$. $P_4/P_T$ is the ratio of the last one by fourth of the frequency range considered (i.e., 1575-2000Hz) and the total power. $dB_{50}$ and $dB_{75}$ indicate the power at 50\%(1000Hz) and 75\%(1500Hz) of 2kHz frequency. $RPF_{50}$ and $RPF_{75}$ denote the ratio of power and frequency at 50\% (1000Hz) and 75\% (1500Hz). $Slope$ features denote the slope between 600Hz and 1200Hz. Baseline feature $P_{4}/P_{t}$ is henceforth referred to as $F_{B}$.

\subsubsection{Proposed features}

Log of the power spectral density has been used to compute the proposed features. \(\boldsymbol{w}\) is calculated in a five-fold cross-validation setup. \(N\) frames before bronchodilator and \(K\) frames after bronchodilator for each subject were separately divided into five folds. eq. 1 was optimized for $\boldsymbol{w}$ by using four-folds data and \(\boldsymbol{w_{opt}}\) are computed. By using \(\boldsymbol{w_{opt}}\), proposed features for the remaining one fold are calculated. For example, if 5 folds are denoted by f1, f2, f3, f4 and f5, then \(\boldsymbol{w_{opt}}\) is learned from f1, f2, f3 and f4 and proposed features are calculated for f5. This is repeated for each fold in a round robin fashion. Hence, five \(\boldsymbol{w_{opt}}\) are learned in such a cross-validation setup and they are used to compute feature vectors in respective folds. Features from all folds are pooled together to examine whether the administering bronchodilator had any effect on the acoustic properties of VBS. The proposed feature is referred to as $F_{P}$. t-test is performed on these pooled features at a significance level of 2\%. 

\paragraph{Frequency band selection using proposed feature}

Frequency bins that contribute more to the discrimination between IPVBS recorded before and after administering bronchodilator, have been calculated by dropping the weights lying below a certain percentile. This experiment has been done on a single weight vector which is calculated by taking an average across all five fold's weight vectors. Steps to calculate features by dropping weights below certain percentile value $\alpha_p$ of absolute weights are shown in eq. 4. Percentile value has been varied from 1 to 99 in a step of 1.
\\
\begin{equation}
\begin{aligned}
\boldsymbol{w_{opt}^a}=[w_{i}]; i=1,2,...171\\
\boldsymbol{w_{opt}^p} =
\begin{cases}
w_i & \text{if}\  \lvert w_i \lvert>\alpha_{p}\\ 
0, & \text{otherwise}
\end{cases}
\end{aligned}
\end{equation}
where $\boldsymbol{w_{opt}^a}$ is obtained by taking average across $w^{opt}$ from five folds, $\boldsymbol{w_{opt}^p}$ is the updated weight vector at the $p^{th}$ percentile, where $\alpha_{p}$ is the $p^{th}$ percentile value of the elements of  $\boldsymbol{w_{opt}}$.

Newly updated weight vector(\(\boldsymbol{w_{opt}^p}\)) has been used to calculate the features which to referred as $F_{\alpha_{p}}$. A statistical test has been performed on this feature set computed from IPVBS recorded before and after administering bronchodilator, respectively. 
For the frequency bins selected across maximum number of subjects, weights have been re-learned using eq. 2. Weights have been re-learned in a subject dependent manner. Analysis of re-learned weights is also presented.

\paragraph{Generalizability of the proposed feature} 
An analysis is carried out to check the generalization of the learned weights across all subjects. For this purposed, a single weight vector has been obtained by performing an average of weight vectors across all subjects. This weight vector has been used to calculate features, denoted by $F_{PA}$, for each subject. Statistical test is carried out to examine if the mean $F_{PA}$ after taking the bronchodilator is significantly different than that before taking the bronchodilator. Similarly, to evaluate the robustness of the learned weights across subjects, weights learned for one subject have been used to compute features for rest of the subjects in a round robin fashion. Features computed in this fashion are referred as $F^{si}$, where $si$ denotes whose subject index which varies from 1 to 30. For example, $F^{1}$ denotes features calculated using weights of subject 1.


\begin{figure}[ht]
	\centering
	\includegraphics[trim={4.3cm 4cm 2 2},scale=.3,clip=true]{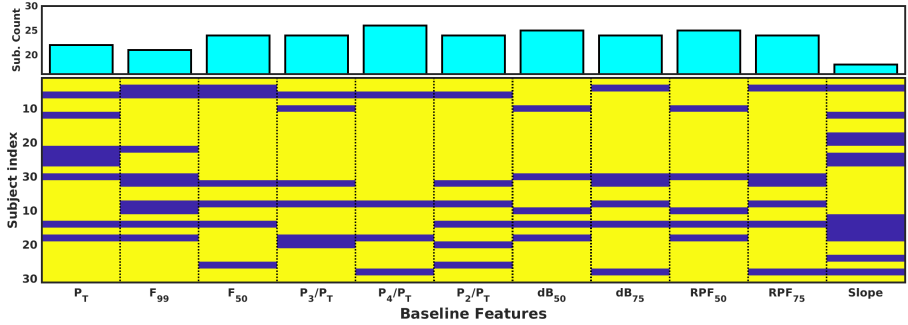}

	\caption{For a given baseline feature on x-axis, yellow color indicates subject index which has shown significant difference between IPVBS recorded before and after administering bronchodilator and blue color indicates a subject that does not have significant difference. Bar graph at the top shows total number of subjects which show significant between differences IPVBS recorded before and after taking bronchodilator using each baseline feature.}
	\label{fig:baseline}
\end{figure}

\begin{figure}
	\centering
	\includegraphics[trim={3cm 1cm 2cm .2cm},scale=.3,clip=true]{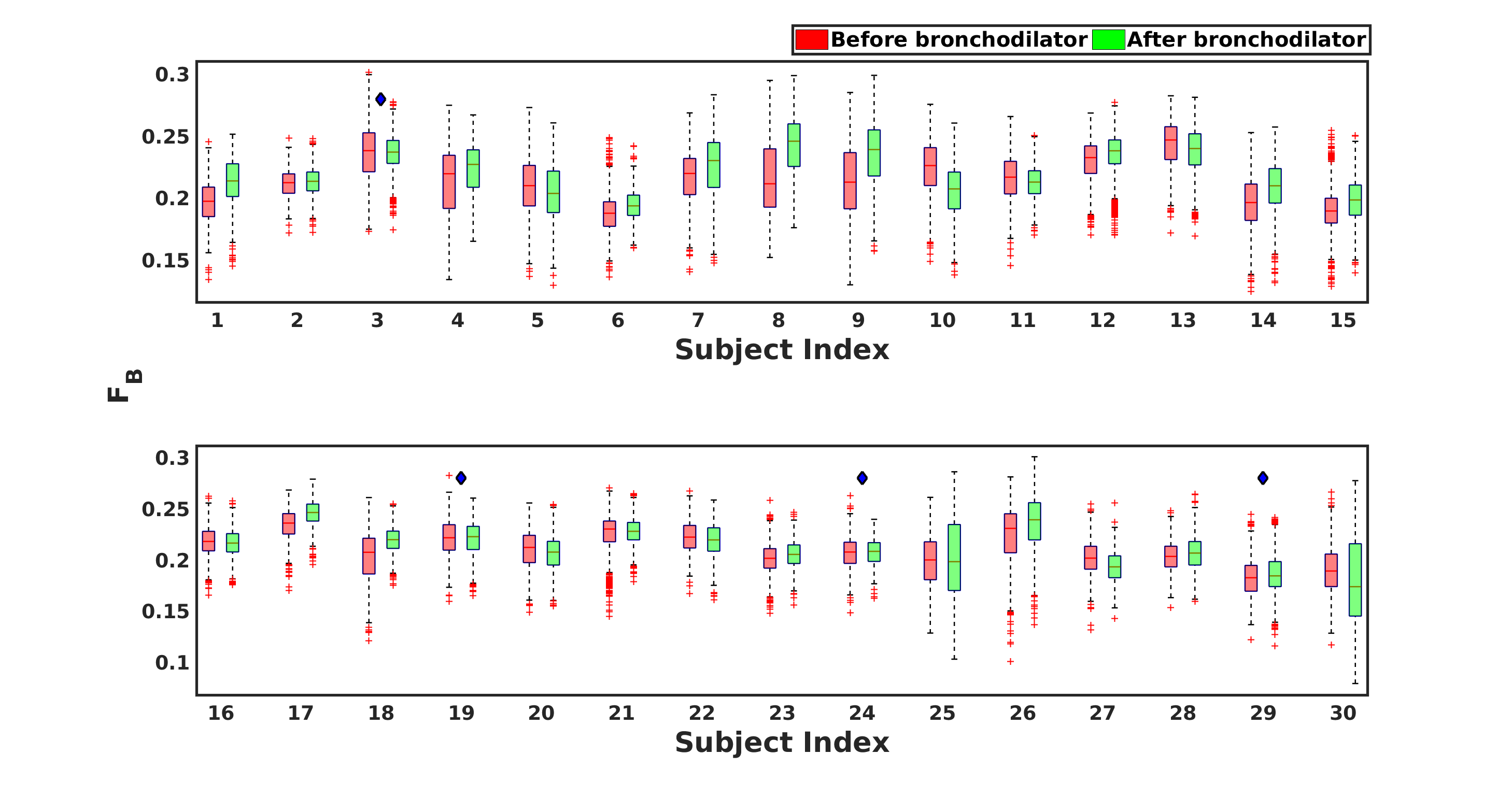}

	\caption{Box plot of best performing baseline feature $F_B$, i.e., $P_{4}$/$P_{t}$, computed on IPVBS recorded before and after administering bronchodilator separately for all the subjects.}
	\label{fig:baseline_best_feature}	
\end{figure}



\section{Results and Discussion}
\subsection{Analysis of baseline features}

Results from the analysis of baseline features are shown in Fig. \ref{fig:baseline}. There are 11 baseline features, which are on the x-axis of Fig. \ref{fig:baseline}.
Subject indices are on the y-axis in Fig. 1. The yellow color for a subject and feature combination indicates that the respective feature is significantly different between IPVBS recorded before and after administering bronchodilator.  The bar graph on the top shows the total number of subjects for whom there is a significant difference between IPVBS recorded before and after administering bronchodilator using each of the eleven baseline features. For example, by using feature $F_{50}$, 2$^{nd}$, 3$^{rd}$, 16$^{th}$, 19$^{th}$, 22$^{nd}$ and 28$^{th}$ subjects do not show significant difference (shown in blue color in Fig. \ref{fig:baseline}). A total of 24 subjects only are found to demonstrate significant difference as shown in the bar graph at the top of Fig. \ref{fig:baseline}.\\

From Fig. \ref{fig:baseline}, we observed that $F_B$ (i.e., $P_4$/$P_T$) performs the best with 26 subjects, whereas the slope from 600-1200Hz performed the worst among all baseline features with only 18 subjects showing significant difference. The best-performing feature $F_B$, indicates that the energy present in the high-frequency region might be useful for the discrimination. The worst performance using the slope over 600-1200Hz could be because it does not capture any information at frequencies higher than 1200Hz.
We also observe that dB$_{50}$ and RPF$_{50}$ perform well with the second-highest number of subjects (25) who demonstrate a significant difference.

Box plot for the best-performing feature $F_B$ is shown in a subject specific manner in Fig. \ref{fig:baseline_best_feature}.
A blue diamond symbol above the box plot in Fig. \ref{fig:baseline_best_feature} indicates subjects who do not show a significant difference in $F_B$ computed on IPVBS recorded before and after administering bronchodilator. In a box plot, top, central and bottom edges indicate $p_{75}$, $p_{50}$, $p_{25}$  which are 75$^{th}$, 50$^{th}$ (median) and 25$^{th}$ percentiles, respectively, of the features values. The whiskers(in black color) extend to the most extreme data points (outliers not included). Data points which are greater than $p_{75}+1.5\times (p_{75}-p_{25})$ and less than $p_{25}-1.5\times (p_{75}-p_{25})$ are treated as outliers as indicated by red color `+' symbol in the box plot. For example, in the case of subject index 6, 75$^{th}$, median and 25$^{th}$ percentiles values are 0.197, 0.187, and 0.177, respectively, before administering bronchodilator (shown in red color box plot in Fig. 2). Maximum and minimum values are 0.249 and 0.136, shown as the extreme whiskers, and a total of 29 outliers are in red, ,`+' symbol. From Fig. \ref{fig:baseline_best_feature}, we can see a change in feature median values before and after administering bronchodilator. However, the sign of change is not consistent across all subjects. 8$^{th}$  subject shows the absolute maximum change of 16.59\% and 24$^{th}$ subject shows an absolute minimum change of 0.3\% in median values from before to after taking bronchodilator among all subjects. 18 subjects have shown an increase, and 12 subjects show a drop in the features median values after taking the bronchodilator. 

\begin{table}[]
	\centering
	\caption{Number of subjects for whom average $F_{P}$, $F_{\alpha_{p}}$ for $\alpha_p$=54 and $F_{PA}$, change significantly after administering bronchodilator compared to before.}
	\label{tab:Weights_1_3exp}
	\begin{tabular}{|c|c|c|c|}
		\hline
		\textbf{\begin{tabular}[c]{@{}c@{}}Feature\\  type\end{tabular}} & \textbf{\begin{tabular}[c]{@{}c@{}}Total significant \\ subjects out of\\ 30 subjects\end{tabular}} \\ \hline
		$F_{P}$                                                                & 30                                                                                                        \\ \hline
			$F_{\alpha_{p}}$                                                               & 30                                                                                                        \\ \hline
		$F_{PA}$                                                                & 26                                                                                                        \\ \hline
	\end{tabular}

%
%

\end{table}


%
%

\begin{figure}[h!]
		\centering
		\includegraphics[trim={0cm 0cm 0 0},scale=.25,clip=true]{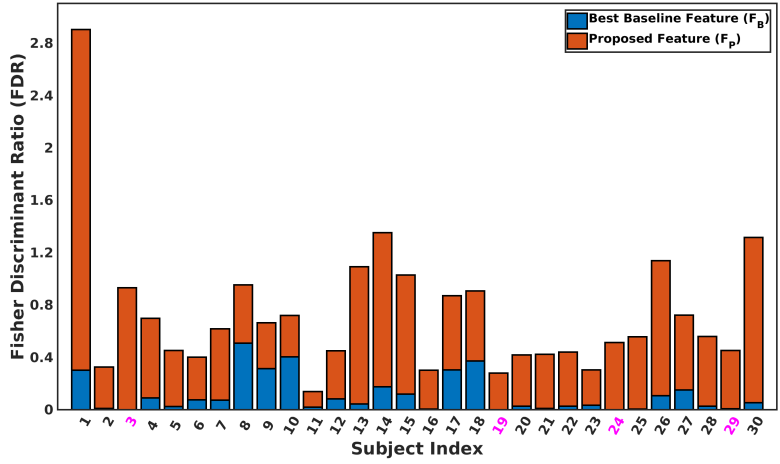}
		\caption{Stacked bar plot of FDR values for best performing feature $F_B$ = $P_{4}$/$P_{t}$ and proposed feature ($F_P$) for all subjects. Indices for the subjects which do not show significant change in $F_{B}$ before and after bronchodilator are shown in magenta color in X-axis.}
		\label{fig:fisher_bsln_prpsd}
\end{figure}
	
	\begin{figure}[h!]
		\hspace{-2cm}
		\includegraphics[trim={3cm 0cm 0cm 2cm},scale=.36,clip=true]{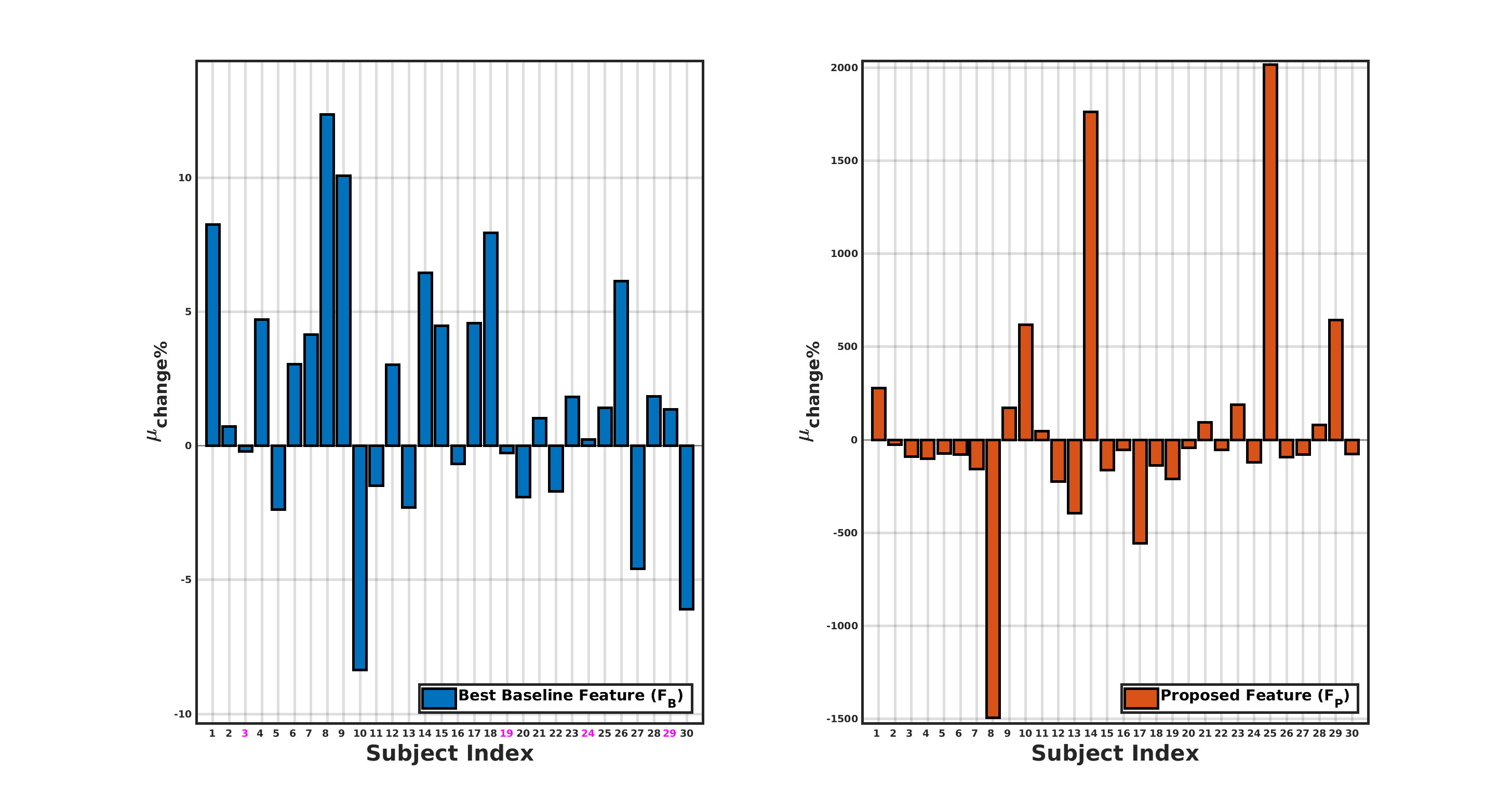}
		\caption{Bar plot of percent change in average value of $F_B$=$P_{4}$/$P_{t}$ and $F_{P}$ before and after administering bronchodilator in separately for every  subject. Indices for the subjects which do not show significant change in $F_{B}$ before and after bronchodilator are shown in magenta color in the X-axis.}
		\label{fig:percent_change_bsln_prpsd}
	\end{figure}
	
	\begin{figure}[h!]
	\centering
	\includegraphics[trim={7.5cm 2cm 0cm 0cm},scale=.75,clip=true]{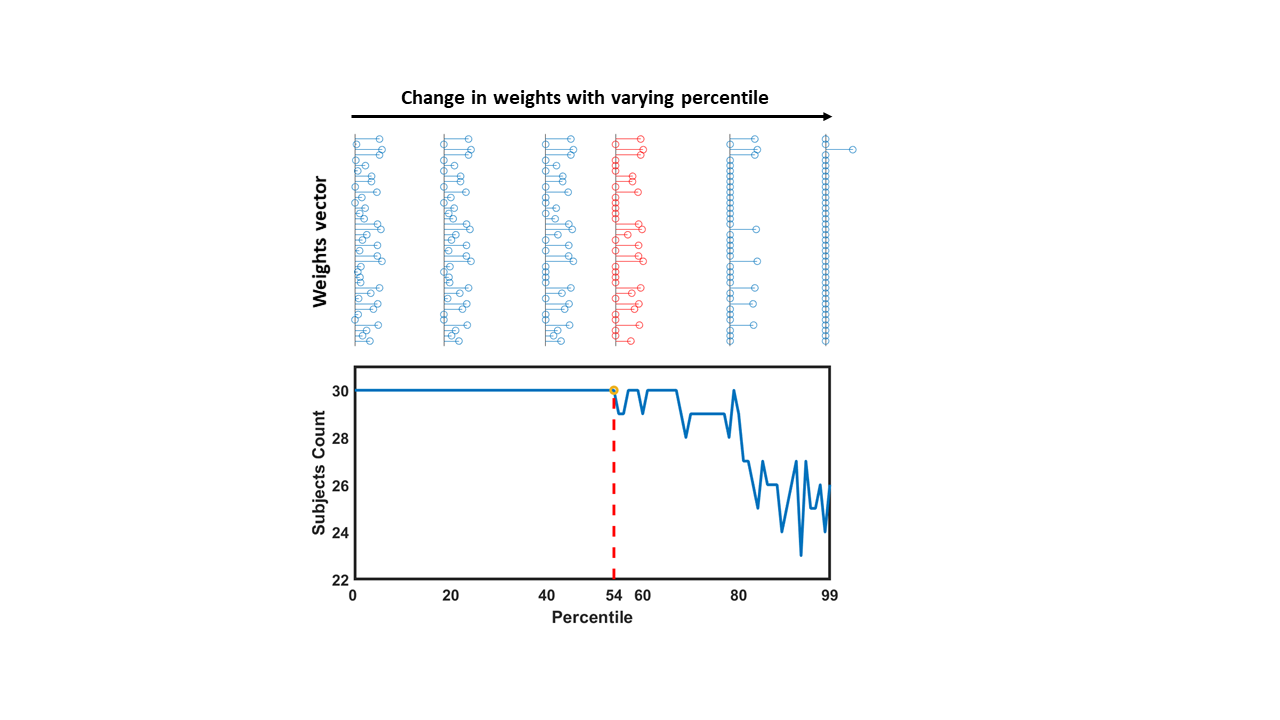}
	\caption{Trend of total number of subjects for whom the proposed feature is significantly different before and after administering bronchodilator with varying weights percentile. Top plot shows an illustrative weight vector when weights are progressively set to zero with increasing percentile value.}
	\label{fig:sub_count_percentile}
\end{figure}

	\begin{figure}[h!]
	\centering
	\includegraphics[trim={3cm 0cm 2cm 1.2cm},scale=.3,clip=true]{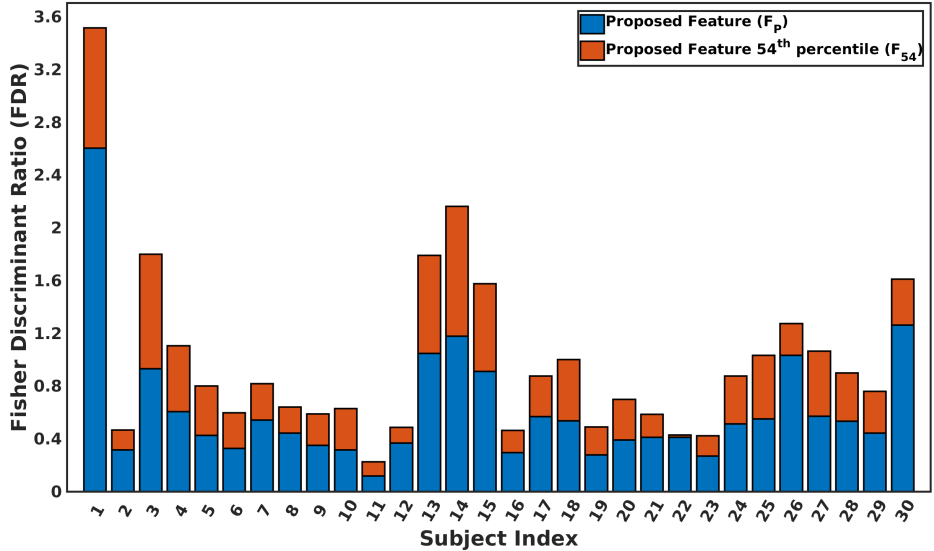}
	\caption{Stacked bar plot of FDR values obtained using  $F_{\alpha_{p}}$ for $\alpha_{p}=54$ and $F_P$ for comparison.}
	\label{fig:fisher_percentile_prpsd}
\end{figure}

\begin{figure}[h!]
	\centering
	\includegraphics[trim={3.4cm 2cm 0cm 0cm},scale=.3,clip=true]{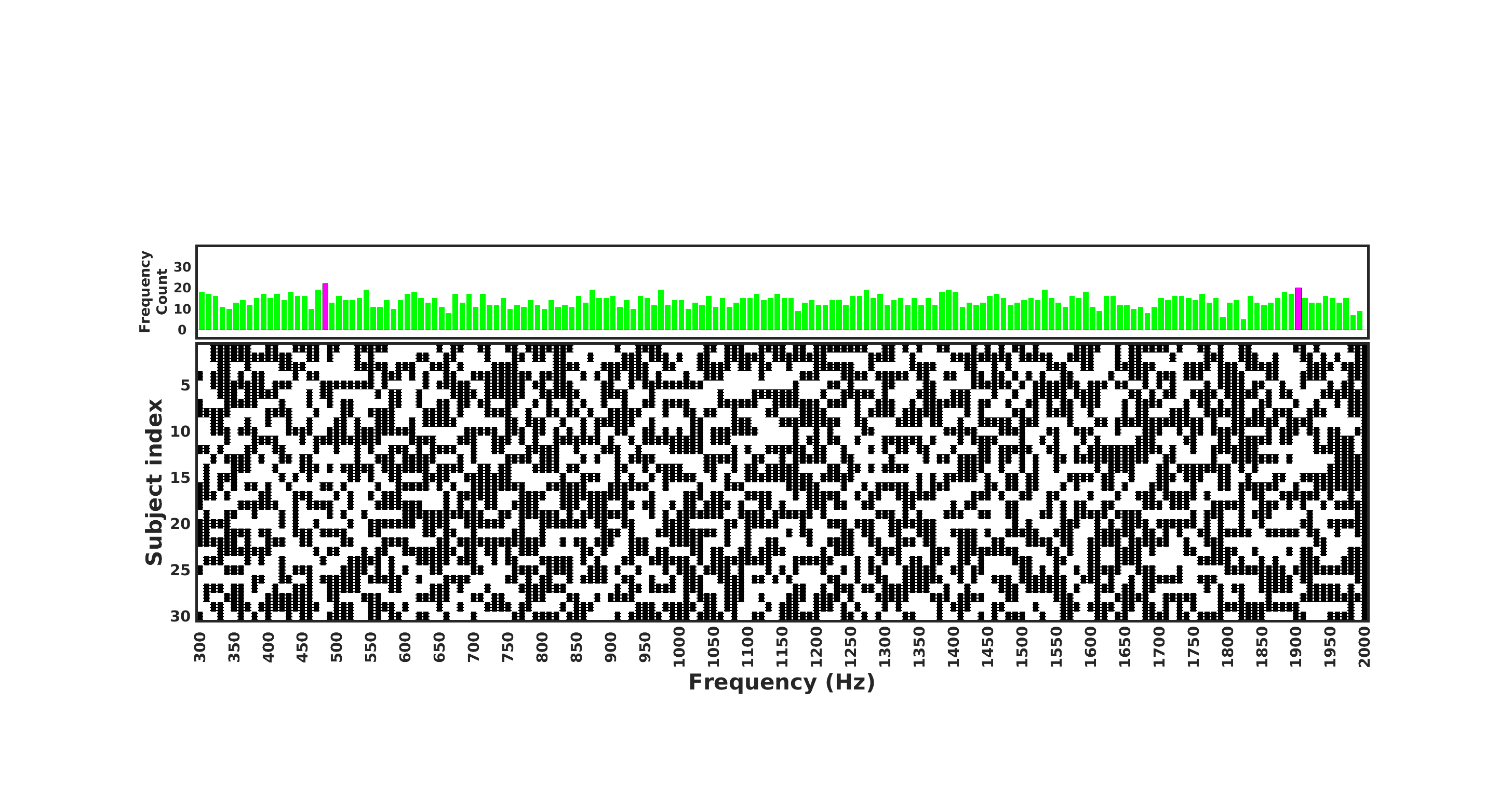}
	\caption{The frequencies which have weights above 54$^{th}$ percentiles in each subject are shown in white. The black dots indicate weights below 54th percentile. Bar plot at the top shows the total number of subjects for whom the weight at a particular frequency is above 54th percentile. Purple color bar in the bar plot shows top two most occurring frequencies.}
	\label{fig:54th_percentile_actual_weights}
\end{figure}


\begin{figure}[h]
	\centering
	\includegraphics[trim={2.2cm 0cm 0cm 0cm},scale=.3,clip=true]{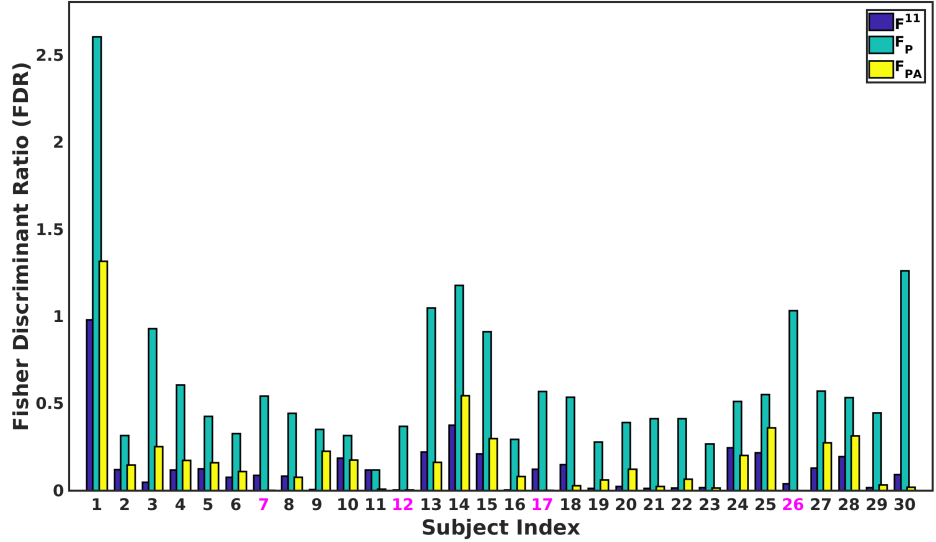}

	\caption{Comparison of FDR by using average weights of all subjects ($F_{PA}$), proposed weights ($F_{P}$) and 11 subject weights ($F^{11}$). By using $F_{PA}$, subjects 7$^{th}$, 12$^{th}$, 17$^{th}$ and 26$^{th}$ (shown in magenta color) do not show significant change pre and post bronchodilator conditions.}
	\label{fig:percent_change_cross_prpsd_avg}
\end{figure}

	\begin{figure}[h!]
	\centering
	 \includegraphics[trim={0cm 0cm 0 0},scale=.3,clip=true]{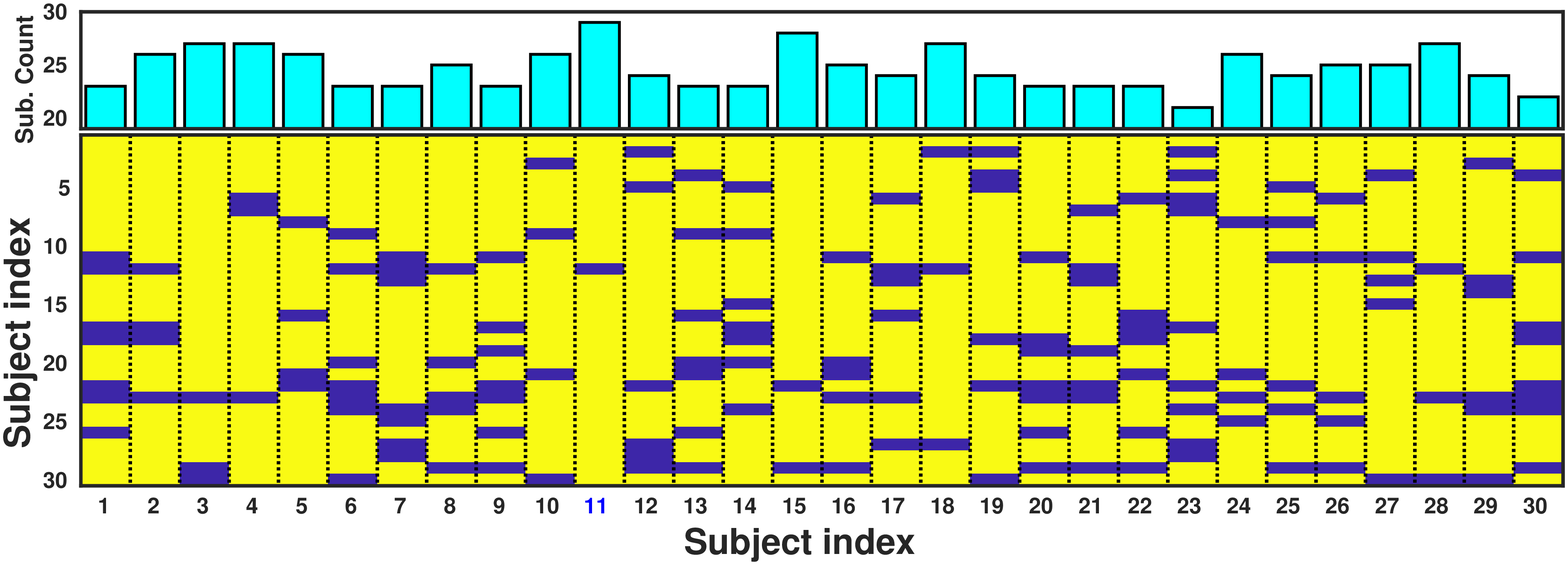}
	\caption{Performance of weights learned for one subjects when used on another subjects. For a given subject weights, yellow color indicates which subjects shows significant difference before and after taking bronchodilator and blue color shows which subject doesn't show significant difference between pre and post bronchodilator conditions.}
	\label{fig:Weights_cross}
	\end{figure}

%

\subsection{Analysis of proposed features}



Table \ref{tab:Weights_1_3exp} shows different kinds of features used, namely, $F_{P}$, $F_{\alpha_{p}}$ and $F_{PA}$ and the corresponding total number of subjects for which the feature was found to be significantly different before and after administering bronchodilator. Proposed features, namely $F_P$ and $F_{\alpha_{p}}$ for $\alpha_p=$54 show a significant difference between IPVBS, recorded before and after administering bronchodilator for all 30 subjects, unlike that of the best baseline feature, $F_B$, which does only for 26 subjects. 

FDR values obtained by using $F_{B}$ and $F_{P}$ for each subject are shown in Fig. \ref{fig:fisher_bsln_prpsd}. The Maximum and minimum FDR calculated using $F_{P}$ are 2.60 and 0.12  for 1$^{st}$ and 11$^{th}$ subjects, respectively. On the other hand, by using $F_{B}$, the maximum and minimum FDR values are 0.51 and 0.003 for 8$^{th}$ and 25$^{th}$ subjects, respectively. These maximum and minimum FDR values using $F_{B}$ are from the 26 out of 30 subjects who showed significant difference before and after administering bronchodilator. 

We observed that inhale sound before and after administering bronchodilator is more separable by using the proposed feature, $F_{P}$, as compared to baseline feature, $F_{B}$, because there is an increase in FDR value using $F_P$ for in all the subjects, except 8$^{th}$ and  10$^{th}$. For the 8$^{th}$ subject, the FDR value is 0.51 by using $F_{B}$ and 0.47 by $F_{P}$. Similarly, these values are 0.40 and 0.32 for the 10$^{th}$ subject by using $F_{B}$ and $F_{P}$, respectively.

Percent change ($\mu{_{change\%}}$) in the mean value of $F_{P}$ and $F_{B}$, before and after administering bronchodilator is shown in Fig. \ref{fig:percent_change_bsln_prpsd} separately for all subjects. Indices of all subjects which do not show significant differences using $F_{B}$ before and after administering bronchodilator are shown in magenta color on the X-axis of Fig. \ref{fig:percent_change_bsln_prpsd}. Interestingly, the sign of the change in the mean is not consistent across subjects. This implies that features do not show a consistent pattern in its change across all subjects after taking bronchodilator compared to before. Similar trend is observed for both $F_B$ and $F_P$. From Fig. \ref{fig:percent_change_bsln_prpsd} we observe that 10 out of 30 subjects show positive change by using $F_{P}$ and 17 out of 26 subjects by using $F_{B}$. The maximum and minimum percent change of 2000\% and 26.38\% are seen by using $F_{P}$ in 25$^{th}$ and 2$^{nd}$ subjects, whereas these are 12.34\% and 0.66\% using $F_{B}$ for 8$^{th}$ and 16$^{th}$ subjects, respectively. Absolute percent change has increased by using $F_{P}$ compared to $F_{B}$ in all the subjects. This suggests that the proposed feature can show more change pre and post bronchodilator as compared to the baseline feature.



%
\subsubsection{Analysis by using $F_{\alpha_{p}}$}
The total number of subjects that have shown significant change after administering bronchodilator compared to before is shown in Fig. 5 with varying weight ($w_{opt}$) percentile. It can be observed from Fig. \ref{fig:sub_count_percentile}, that after 54$^{th}$ percentile ($\alpha_p$=54), discrimination between features computed on IPVBS recorded before and after administering bronchodilator reduces. This leads to a reduction in the number of subjects for whom a significant difference is observed. It is worth mentioning from this experiment that the weights above 54$^{th}$ percentile are enough to significantly discriminate IPVBS before and after administering bronchodilator in all the subjects. These findings suggest that all frequencies in the range of 300-2000Hz may not be required for discrimination.

Features $F_{\alpha_p}$ corresponding to $\alpha_p$=54 referred as $F_{54}$.
FDR value is used to compare the performance using $F_{54}$ and $F_P$ and it is shown in Fig. \ref{fig:fisher_percentile_prpsd}. We observe that for all 30 subjects, there is a decrease in FDR value by using $F_{54}$ as compared to that using $F_{P}$. 


Maximum drop of 1.69 in FDR occurred for the 1$^{st}$ subject when $F_{54}$ is used. Maximum FDR using $F_{54}$ is 0.98, which is less than the maximum FDR using $F_P$ but greater than the maximum FDR value obtained using the baseline features $F_{B}$. This suggests that optimally chosen frequency bins corresponding to $F_{54}$ provide more discrimination than baseline features.

\subsubsection{Frequency band selection by using $F_{\alpha_{p}}$ for $\alpha_{p}$=54}

Our next goal is to find out frequency bands that contribute maximally to the discrimination of the proposed features computed before and after administering bronchodilator. For this experiment, frequency bins which are common across majority of the subjects are selected, after weights below 54$^{th}$ percentile are assigned to zero. Frequencies bins which have non-zero weights for each subject in the frequency range of 300-2000Hz are shown in Fig. \ref{fig:54th_percentile_actual_weights}. Black and white color dots indicate zero and non-zero weights, respectively, for a subject. The bar graph at the top in Fig. \ref{fig:54th_percentile_actual_weights} shows the total number of subjects for whom the weight was non-zero for the corresponding frequency bin.
Two frequency bins, namely 480Hz and 1900Hz, were found to be common across 22 and 20 subjects, respectively, as shown in purple color in the bar plot of Fig. \ref{fig:54th_percentile_actual_weights}.

Frequency bin-specific weights had been re-learned to analyze the role of frequency bins on proposed features' performance. Out of the total 171 bin, only bins corresponding to 480Hz and 1900Hz, have been used to learn the weights. Features calculated only with these two frequency bins using re-learned weights showed significant differences between IPVBS recorded before and after administering bronchodilator in 24 subjects. This shows that, while two frequency bins, namely 480Hz and 1900Hz, provide discrimination for most of the subjects, they, unlike $F_P$, do not capture entire spectral characteristics that are required to achieve significant discrimination for all 30 subjects.

Further experimentation by using grid search was done to find the frequency bins which corresponds to frequency range as follows:
\begin{equation}
480-\alpha\leq f_1 \leq 480+\beta, \quad 1900-\gamma\leq f_2 \leq 1900+\delta, \quad f_1<f_2
\end{equation}
where, $0\leq\alpha\leq180$, $0\leq\beta\leq1420$, $0\leq\gamma\leq1420$ and  $0\leq\delta\leq100$. Grid search is done with a step size of 20Hz. Weights corresponding to bins in the frequency range, $f_1$ and $f_2$ are re-learned for each combination of \(\alpha\), \(\beta\), \(\gamma\) \text{and} \(\delta\) values. The best choices, \(\alpha^*, \beta^*, \gamma^* \text{ and } \delta^*\) are selected for which the maximum total number of subjects showed significant differences between IPVBS recorded before and after administering bronchodilator. When for multiple combinations of $\alpha$, $\beta$, $\gamma$ and $\delta$ resulted in the highest number of subjects, the best choice is made for which the mean of $\mu_{change}$\% across all subjects is the maximum. From grid search, we have found that \(\alpha^*=80, \beta^*=20, \gamma^*=420  \text{ and } \delta^*=0\) and all 30 subjects showed significant difference before and after bronchodilator. These values of \(\alpha^*, \beta^*, \gamma^* \text{ and } \delta^*\) lead to \(400\leq f_1\leq 500\) and \(1480\leq f_2 \leq 1900\). Through this experiment it is clear that, high frequency region of VBS is more sensitive to the bronchodilator treatment. These findings are in an agreement with Tabata et al. \cite{tabata2018changes}.

%



\subsubsection{Results by using $F_{PA}$}

In order to investigate the generalizability of the learned weights across subjects, the mean of all learned weights across subjects in all folds is calculated (denoted by $F_{PA}$). Using $F_{PA}$ 26 subjects have shown significant discrimination between pre and post bronchodilator conditions as given in Table \ref{tab:Weights_1_3exp}. Comparison of FDR values using $F_{PA}$ and $F_{P}$ are shown in Fig. \ref{fig:percent_change_cross_prpsd_avg}. FDR is nearly zero for subjects 7$^{th}$, 12$^{th}$, 17$^{th}$ and 26$^{th}$ using $F_{PA}$. Minimum and maximum FDR values are 0.009 and 1.317 for 11$^{th}$ and 1$^{st}$ subject by using $F_{PA}$ which is better than the $F_B$, whereas by using $F_{P}$, FDR values for the same subjects are 0.118 and 2.604, respectively. FDR is higher across all the subjects by using $F_{P}$ compared to $F_{PA}$.
The poor performance of $F_{PA}$ indicates that, it is unable to capture all kinds of variability like gender, age, asthma severity, which are speaker specific. On the other hand, when we use subject-specific weights, $F_{P}$ does capture speaker-specific traits.

Subject-specific weights are used on other subjects to find the features, in order to examine the robustness of the learned weights across subjects. The findings from this experiment are given in Fig. \ref{fig:Weights_cross}. From Fig. \ref{fig:Weights_cross} it can be seen that 11$^{th}$ subject (referred as 11 in the Fig. \ref{fig:Weights_cross}) performed the best, as it can discriminate 29 subjects.
Performance of 11$^{th}$ subject's, weights is even better than the $F_{PA}$, which discriminates IPVBS only for 26 subjects. Performance using the weights of the 23$^{rd}$ subject is the worst among all subjects as it can significantly discriminate IPVBS for only 21 subjects.

Comparsion of FDR values among $F_{PA}$, $F_{P}$ and $F^{11}$( $F^{11}$ denotes features calculated by using 11$^{th}$ subject weights) are shown in Fig. \ref{fig:percent_change_cross_prpsd_avg}. From Fig. \ref{fig:percent_change_cross_prpsd_avg}, we observe that $F_{P}$ performs better than $F_{PA}$ and $F^{11}$. $F^{11}$ does not differentiate 12$^{th}$ subject significantly just like $F_{PA}$ as both the features have low FDR for this subject.
Another finding is that the features calculated by using the weights of 11$^{th}$ subject have shown significant discrimination between pre and post bronchodilator conditions in 7$^{th}$, 12$^{th}$, 17$^{th}$ and 26$^{th}$ subjects, which is not the case with $F_{PA}$. FDR values also improve in these 3 subjects by using 11$^{th}$ subject's weights.

\section{Conclusion}

The study presented in this paper analyzes the pre and post bronchodilator effect on the inhalation sound recorded at the mouth. This study uses 30 asthmatic subjects. It has been observed that subject specific features performed better than subject independent features to discriminate pre and post bronchodilator condition. Two different types of features are used in this study, namely, proposed features and baseline features. Proposed features are found to be better to discriminate pre and post bronchodilator conditions as compared to the baseline features in each of the 30 subjects. Experiments using the proposed features reveal that 400-500Hz and 1480-1900Hz frequency bands provide sufficient information to discriminate pre and post bronchodilator conditions in each of the 30 subjects. Findings of this work suggests that, inhalation sound recorded at mouth can be a good stimulus to discriminate pre and post bronchodilator conditions in asthmatic subjects. Future work includes analysis of exhale as well as full breath signal recorded at mouth for the similar tasks. Even with more data, data-driven approaches like neural networks can be used to learn better representations for the discrimination between pre and post bronchodilator conditions.

\section*{Conflict of interest}
None declared

\bibliographystyle{unsrt}
\bibliography{ms}

\end{document}